*Love Thy Neighbor? Perceived Community Abidance and Private Compliance to COVID-19 Norms in India*

Upasak Das
*Global Development Institute, University of Manchester*
*upasak.das@manchester.ac.uk*

Prasenjit Sarkhel
*Department of Economics, University of Kalyani*
*prasenjitsarkhel@gmail.com*

Sania Ashraf
*Center for Social Norms and Behavior Dynamics*, University of Pennsylvania
*asania@upenn.edu*

*Abstract:* To tackle the COVID-19 infection, frequent hand-washing and respiratory hygiene measures like wearing face-masks and maintaining a minimum distance from others are often considered as the most important ways to prevent the spread of the virus. These non-pharmaceutical interventions, in the absence of vaccines or pharmaceutical therapies, are expected to reduce the infection rate over time. These behaviors are often argued to be pro-social, which are driven by multiple social motives such as, social norms, reciprocity, cooperation and social learning. However one must incur private cost, which may be due to reasons like economic or fatigue to benefit or protect others. Using self-reported data across India (n=934) collected through social media platforms, we assess if changes in perceived community compliance can predict changes in individual compliance behavior, controlling for the potential confounders. We observe statistically significant and positive relationship between the two, even after accounting for omitted variable bias and other robustness and falsification tests, plausibly allowing us to view the results from a plausible causal lens. Further, we find subsequent lockdowns such as the ones imposed in India, having a detrimental effect on individual compliance though the gains from higher perceived community compliance seem to offset this loss. We also find that sensitization through community can be particularly effective for people with pre-existing co-morbidities. Our findings underscore the need for multi-level behavioral interventions involving local actors and community institutions to sustain private compliance during the pandemic. We suggest these interventions need to be specially targeted for individuals with chronic ailments especially in places like medical shops and health centres where they are more likely to visit. Apart from the policy implications, the paper assumes contextual importance as India remains as among the most affected countries across the globe.

Introduction

The novel coronavirus (COVID-19) pandemic has spread globally claiming more than a million lives.[1] To tackle it, at the individual level, the World Health Organization (WHO) and other experts recommend frequent hand-washing and respiratory hygiene by wearing a mask and maintaining a minimum distance from others as the most important ways to prevent the spread of the virus.[2] These non-pharmaceutical interventions (NPI), in the absence of vaccines or pharmaceutical therapies, are expected to reduce the infection rate over time. In the initial period, government across developed and developing countries relied on deterrence measure like lockdown to induce compliance. These measures involved restrictions on human movement and accompanying closure of public institutions like schools and colleges as well as work places and business institutions. Given the fact that these are costly to adopt in terms of time and effort, when imposed uniformly on a heterogeneous population of varying income and social class, the private level of compliance would vary. Some have argued that those with high opportunity cost would then exercise lesser precaution, particularly if long drawn movement restrictions impinge upon their income and livelihood. Additionally, as the economy reopens the need to adopt alternative interventions for inducing optimal social distancing becomes an important policy concern.

---
[1] Coronavirus COVID-19 Global Cases by the Center for Systems Science and Engineering (CSSE) at Johns Hopkins University (JHU). https://coronavirus.jhu.edu/map.html (accessed on October 14, 2020)

In as much as the compliance to COVID-19 protocols confers collective benefits, substantial concerns include the associated private cost such as inability to work, or poorer wellbeing due to social isolation and the disruption of their livelihood (Ahmed et al. 2020; Lancet, 2020). As a result, voluntary compliance effort might fall short of the optimum over time. Some of the compliance activities are costly and difficult to monitor like use of sanitizers and hand-washing. Thus, imposing penalties or sanctions for partially observable compliance behavior, may not guarantee best outcomes particularly if violation of COVID-19 related norm can take place in the private domain. It is here that the importance of community arises because much of the motivation for compliance behaviors are pro-social, which can be driven by multiple social motives such as, social norms, reciprocity, cooperation and social learning (Batson and Powell, 2003; Dovidio et al. 2017). Studies have shown that people are strongly motivated by their peers, through peer pressure to align with the group and to reciprocate pro-social behavior such as wearing a mask or staying indoors for the greater public good (Rand et al. 2014). Understanding these motivations during the COVID-19 pandemic is important to inform effective public messaging efforts. In this context, using online self-reported data collected from

---

[2] World Health Organization (WHO). Coronavirus disease (COVID-19) advice for the public. https://www.who.int/emergencies/diseases/novel-coronavirus-2019/advice-for-public. (accessed on October 14, 2020)

934 respondents across India, we assess if positive perception about community compliance behavior is associated with higher levels of individual abidance of the preventive measures. Further, we also examine the changes in compliance behavior over the subsequent lockdowns implemented in India from March 25 to May 31, 2020 after the outbreak and assess the implications of better perceived community compliance. Since individuals with chronic co-morbidities are more susceptible to serious illness, we also assess the implications of community abidance over these set of individuals as well.

The importance of the context is significant here since India has been among the most affected countries in the world. Currently (October 11, 2020),, more than 7.1 million infected cases have been reported with a death toll of more than 110,000 . As a preventive measure, the government had ordered a nationwide massive lockdown in four phases: Phase 1: 24 March- April 14; Phase 2: April 15 to May 3; Phase 3: May 4 to May 17 and Phase 4: May 18 till May 31. In all these phases, India imposed the strictest lockdown rules curbing individual movements for non-essential purpose, mandating face-masks and allowing no social gatherings. Despite this we find a substantial increase in the number of infection cases over these four lockdowns (Figure1). Notably, the economic loss during this period was substantial as India recorded a slump by about -23.9 percent (year-on-year) in the quarter of April to June, 2020.[3] Studies have documented the huge economic disruptions among different section in the economy in India (Ceballos et al. 2020; Mehra et al., 2020)

---

[3] https://economictimes.indiatimes.com/news/economy/indicators/gdp-growth-at-23-9-in-q1-worst-economic-contraction-on-record/articleshow/77851891.cms?from=mdr (accessed on October 14, 2020)

Figure 1: State-wise COVID-19 cases per-million population in India over the four phases of

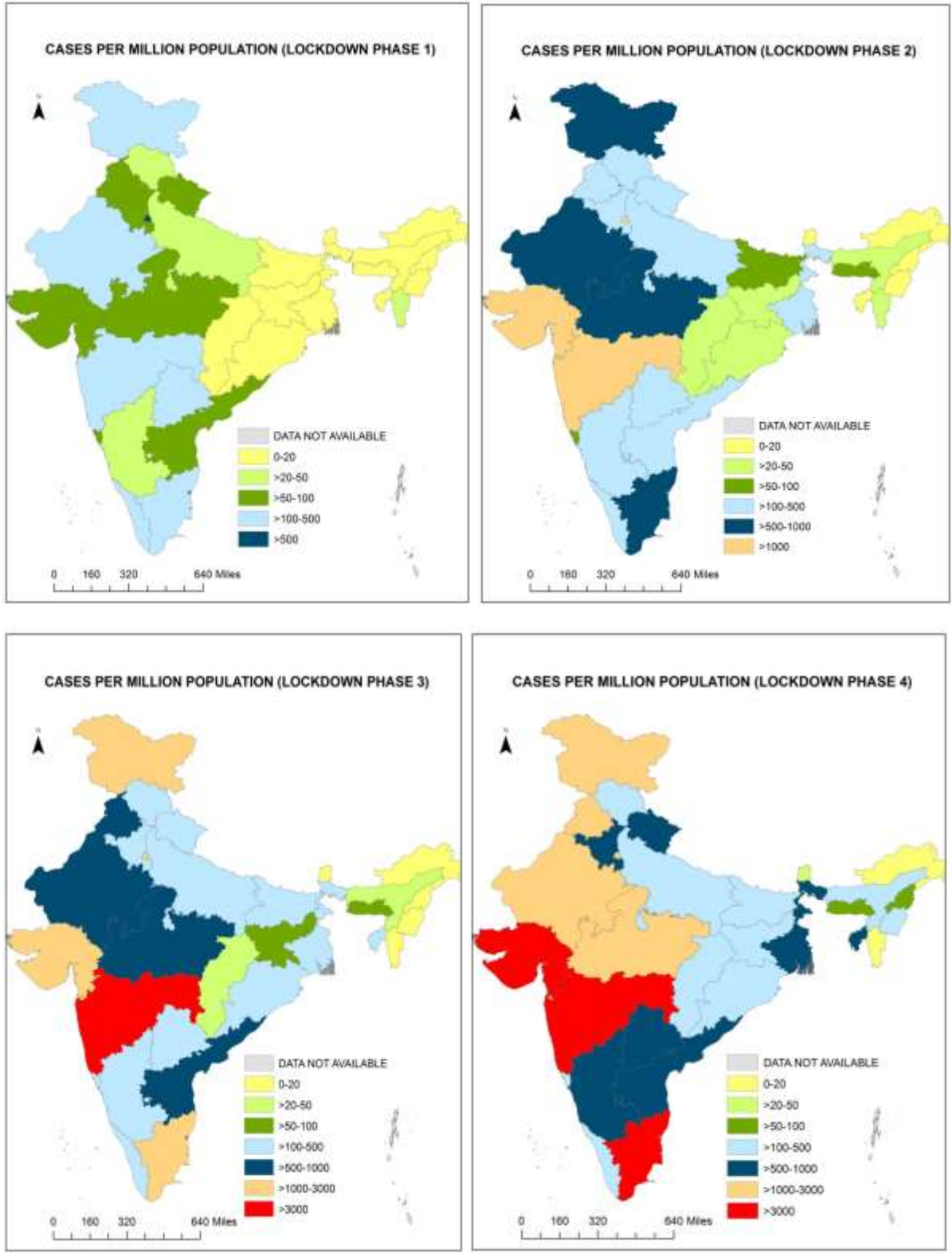

Source: The Socioeconomic High-resolution Rural-Urban Geographic Dataset on India (SHRUG). http://www.devdatalab.org/shrug

Theoretically, one major driver for private compliance would be to ensure better health and avoid the possibility of getting infected. The other driver that confers collective benefits, stems from moral motivation to protect others. Variants of moral motivation like pure altruism (Andreoni, 1988; Andreoni, 1989), and improved self-image (Brekke et.al., 2003) have been used to explain voluntary contribution in public good contexts (Chorus, 2015). In this strand of literature, individual adherence to pro social behavior occurs because a positive utility is derived from matching actions that are considered ideal. In fact, studies have argued how individual behavior are often driven by norms, social expectations and first order beliefs about whether her community follows a particular behavior (Lyon, 2000; Cialdini and Goldstein, 2004; Bicchieri, 2017). In the context of disease outbreaks like COVID-19 there is a possibility that abidance to the compliance protocols might be motivated by such pro social concerns. On the other hand, economic incentives often interact with moral incentive such that collective behavior may impose a decision externality on private action (Collier and Venables, 2014). Better community preparedness that pose lesser chances of infection may induce individuals to reduce her preventive efforts. Thus, the net effect of community compliance level on individual action is ambiguous[4]. Moreover, government reliance on lockdowns may intensify economic hardship which could weaken the community ties that motivates compliance. In fact, studies have documented that individual compliance might be inversely related with the duration of social distancing measures (Briscese et. al., 2020; Moraes, 2020). While economic hardship and violation of COVID-19 related norms have received considerable attention in the current scenario, the impact of moral motivation and health seeking behavior is relatively less explored. Our study adds to this evidence by analyzing the direction of the effect of community compliance on individual protective action.

The findings indicate a statistically significant, robust and positive relationship between perceived community compliance and individual compliance levels. Notably, these positive relationships hold even after accounting for the potential unobservable that include strategies developed by Altonji et al. (2005) and Oster (2019) indicate that our findings might be viewed through the causal lens though concerns about reverse causality remain. We also find compliance for co-morbid respondents to increase systematically with higher levels of perceived community compliance. Interestingly we find a significant decline in compliance levels with subsequent lockdowns but the gains through higher community compliance seems to offset this loss. Given this evidence, we propose a set of recommendations that can increase adoption of preventative measures to curb COVID-19 transmission. Among others, we propose behavioral interventions and outreach programmes through active local community institutions that makes community compliance visible and salient among its members. Further underestimation of the actual community compliance can be corrected since it can reinforce positive norms.

---

[4] Prolonged self-isolation when the outside community is engaging in proper preventive measures might adversely affect physical and mental health (Rozenkrantz et al. 2020).

The structure of the paper is as follows. Section 2 gives a brief outline of the theoretical underpinnings and conceptual framework. Section 3 discusses the data and the variables. Section 4 and section 5 reports the results while the associated discussion are presented in Section 6. A brief conclusion is provided in section 7.

**2. Conceptual Framework**

Consider a society with *N* individuals and the utility function of individual, *i* is given by

$$U_i = U(x_i, h_i, I_i)..(1)$$

Here $x_i$ is the consumption of private goods $h_i$ denotes the health status and $I_i$ is the measure of individual's self-image. We assume that the utility function is continuous and strictly quasi-concave. We suppose further that the health stock of the individual depends on his compliance level, $c_i$, say that is given by the amount of time, she distances herself from the society and/or the amount of time that she engages in protective activities like wearing mask and hand-washing among others. However, private health stock would also depend on the extent to which community in his neighborhood follows the health norm. Given the infectious nature of the COVID-19, we presume community behavior would have an impact upon individual health through contagion dynamics. For simplicity, we assume that the individual interacts only within her own neighborhood, which, given the extended lockdown period and high neighborhood interaction seems plausible in the Indian context. Let the perception about community health norm be given as $\bar{c}_i$. We then suggest the following functional form of the health status of the individual using

$$h_i = h(c_i, \bar{c}_i) \ldots \ldots (2)$$

Thus, health is a concave function of individual compliance.

The other role that community compliance has is through its moral appeal. We assume that individual converge to behavioral conformity by attempting to adopt the community level compliance. In the literature such pro-social moves have been modeled as one emanating from stigma from not following socially prescribed action (Akerlof and Kranton,2000; Brekke et.al., 2003) or individual might want to do what the others are doing in order to signal themselves as desirable social types (Bernheim,1994; Bigenho and Martinez, 2019). Following these literature, we assume that the representative individual would attempt to bolster his self-image by following the compliance behavior of the community

Accordingly, we define the self-image function

$$I_i = -\alpha(c_i - \bar{c}_i)^2, \alpha > 0 \ldots \ldots (3)$$

as in Brekke et al. (2003). The function is concave in $c_i$, $I_i$ is increasing if $c_i < \bar{c}$, it is zero if $c_i = \bar{c}_i$ and it is decreasing if $c_i > \bar{c}_i$.

The time constraint of each individual is

$$c_i + l_i = T \ldots \ldots (4)$$

where $l_i$ is the time allocated for work. Assuming that hourly wage rate is $w$ we can replace $x_i = w(T - c_i)$ in the utility function, that is $x_i$ is the numeraire good. The individual maximizes (1) subject to his health production function, self-image function and time constraint in (2) to (4). The first order condition gives

$$c_i = -\frac{w}{2}\frac{U_x}{aU_I} + \bar{c}_i + \frac{1}{2}\frac{U_h}{aU_I}\frac{\partial h}{\partial c_i} \ldots \ldots (5)$$

Equation (5) defines the relation between $c_i$ and $\bar{c}_i$. The first term on the RHS denotes the opportunity cost of consumption relative to self-image from changes in individual compliance. Since $U_x, U_I > 0$ this term is negative for a given wage. On the other hand, the third term on the right hand side gives the marginal health benefit relative to self-image from changes in individual health compliance and given the concavity assumption is positive. Thus, the extent to which individual compliance would deviate from the community compliance would depend on the relative strength of these two terms.[5]

**3. Data, variables and estimation**

*3.1 Data*

The data used in the paper is obtained from an online survey of 934 respondents across India from April 27 to May 31, 2020 mainly through snowball sampling. Since this survey was conducted during the lockdown face to face interviews or even phone call interviews could not be conducted in absence of a representative set of phone numbers. The survey was distributed via online platforms like the WhatsApp, Facebook and Twitter among others.

The survey was administered in four main languages: Bengali, Hindi, English and Tamil and hence our survey has majority of the respondents from West Bengal, Tamil Nadu and parts of the Hindi speaking belt that includes the Northern India. The questionnaire form and the data sheet were administered in Google form. Research assistants who were native speakers of these languages pre-tested the questionnaires. Suggestions to improve comprehension and flow of the survey were incorporated. We used several channels of distribution to ensure variation in our sample with respect education, occupation, age and gender. This study was approved by the Institutional Review Board at the University of Pennsylvania (Protocol No.: 843016).

The survey focused on individual level compliance and gathered data on the respondent's behavior a week prior to the survey on a set of compliance protocols. These protocols included (i) whether the respondent used face-masks while going out or speaking to others outside their household members; (ii) whether she maintained six-feet distance in public; (iii) whether she washed hands with soap more than that before the lockdown started and (iv) whether she used hand sanitizer. Further, it collected information on perceptions about the community compliance through a number of indicators. The related questions are: out of 10 individuals in the community, how many does s/he think have (i) stepped outside for non-essential purpose; (ii) maintained 6 feet of distance while talking and (iii) washed their hands regularly with soap. The survey also collected information on respondent's belief about their community on indicators that include questions like: out of 10 individuals in the community, how many does she think have believe that (i) people should step out only for essential purposes; and (ii) at least 6 feet/ 2 meters distance should be maintained while talking. In addition to these questions, we collected information on other household and individual level characteristics. Further details have been included in table 1.

---
[5] An exploration with functional form of a utility function is given in Appendix 1.

Table 1: Survey Instruments

| Variables | Options |
| --- | --- |
| *Individual Compliance* | |
| How frequently did you do the following in the last week? [I maintained at least 6 feet/ two meters in public] | 1. Always |
| | 2. Most of the time |
| How frequently did you do the following in the last week? [I used a mask when I went out] | 3. Half of the time |
| | 4. Rarely |
| | 5. None of the time |
| How frequently did you do the following in the last week? [I washed my hands more frequently] | |
| How frequently did you do the following in the last week? [I used a hand sanitizer] | |
| *Perception about Community compliance* | |
| Out of 10 people in your community, how many do you think have stepped out of their house for non-essential purposes? | 0-10 |
| Out of 10 people in your community, how many do you think maintain at least 6 feet/ 2 metres while talking? | |

*Beliefs about community compliance*

Out of 10 people in your community, how many do you think believe that people should step out only for essential purposes?

Out of 10 people in your community, how many do you think believe least 6 feet/ 2 meters distance should be maintained while talking?

*Anxiety*

| To what extent do the following statements apply to you? [I am worried about my family's health] | 1 Doesn't apply |
| To what extent do the following statements apply to you? [I am anxious all the time] | 2 Somewhat does not applies |
| To what extent do the following statements apply to you? [My and my family members' jobs are secure in this crisis] | 3 Neither |
| | 4 Somewhat applies |
| To what extent do the following statements apply to you? [I am worried about my finances because of this crisis] | 5 Strongly applies |

*3.2 Variables and estimation*

We use the first two indicators related to stepping outside for non-essential purpose and maintaining six feet distance while interacting with others for calculating the perception about community compliance. We left out the third indicator pertaining to hand-washing since this behavior about the others in the community is not observable for the respondents. We standardized these variables and use that to calculate perception about community compliance. Our dependent variable is individual level compliance, which is calculated by standardizing the four protocol adherence variables that we mentioned. We use Ordinary Least Squares (OLS) to regress the individual level compliance score on the perception about community compliance scores controlling for the potential confounders. The list of confounders and the associated variable construction are given in table 2.

Table 2: Description of the variables

| Variable Name | Description | Type of variable |
|---|---|---|
| *Outcome Variable* | | |
| Individual compliance score | We use the four indicators of individual compliance to calculate the score. First we created dummy variables to indicate whether the respondent complied all the time in the last week or not<br><br>We first normalize each of the four dummy variables to have zero mean and a standard deviation of one. Then we sum up the standardized measures and then again re-standardize such that the final score would have 0 mean and one standard deviation. This method has been used by Heath and Tan (2019) for their calculation of women autonomy. | Continuous |
| *Main variable of interest* | | |
| Community compliance score | We use the two indicators capturing perception about community compliance to calculate this score. Like in the earlier case, we first normalize each of the two indicators to have zero mean and a standard deviation of one. Then we sum up the standardized measures and then again re-standardize such that the final score would have 0 mean and one standard deviation. | Continuous |
| *Control variables* | | |
| Beliefs about community score | We use the two indicators capturing beliefs about community compliance to calculate this score. Like in the earlier case, we first normalize each of the two indicators to have zero mean and a standard deviation of one. Then we sum up the standardized measures and then again re-standardize such that the final score would have 0 mean and one standard deviation. Here, for the second indicator (Out of 10 people in your community, how many do you think have stepped out of their house for non-essential purposes?), we used subtracted 10 from the responses and then took the modulus value since both indicators would now present positive beliefs about the community. | Continuous |
| Lockdown dummy | Two categories:<br><br>1: Phase 2 lockdown<br>2: Phase3/ Phase 4 lockdown | Categorical |

| Age category | Five categories: | Categorical |
|---|---|---|
| | 1: Below 25 years<br>2: 26 to 30 years<br>3: 31 to 40 years<br>4: 41 to 50 years<br>5: Above 50 years | |
| Sex | Two categories: | Categorical |
| | 1: Female<br>2: Male/ cannot say (Reference group) | |
| Marital Status | Two categories:<br>1: Single<br>2: Cohabiting/ Married/ Others | Categorical |
| Education | Four categories: | Categorical |
| | 1: No formal education/ 10 years or below 10 years of education<br>2: Higher secondary/ College education<br>3. Graduate<br>4: Post graduate or above | |
| Government job | Two categories: | Categorical |
| | 1: Engaged as a regular/ contractual worker in government sector<br>2: Does not work/ Home-maker/ Housewife/ Student/ Farmer business owner/ Daily wage earner/ Engaged as a regular/ contractual worker in private sector | |
| Contractual job | Two categories: | Categorical |
| | 1: Engaged as a contractual worker in government or private sector<br>2: Does not work/ Home-maker/ Housewife/ Student/ Farmer business owner/ Daily wage earner/ Engaged as a regular worker in government or private sector | |
| Large family | Two categories: | Categorical |
| | 1: Four members or less in the household<br>2: Five or more members in the household | |
| Number of elderly | Three categories: | Categorical |
| | 1: No elderly in the household in the age group 65 years and above<br>2: One elderly in the household in the age group 65 years and above | |
| | 3: More than one elderly in the household in the age group 65 years and above | |
| Number of rooms | Three categories:<br><br>1: One or two rooms<br>2: Three rooms<br>3: More than three rooms | Categorical |
| Owned house | Two categories:<br><br>1: Lives in an owned standalone house<br>2: Lives in a rented house or apartment complex | Categorical |
| Co-morbidity | Two categories:<br><br>1: Do not suffer from major ailments<br>2: Suffers from either high blood pressure or diabetes | Categorical |
| Family co-morbidity | Two categories:<br><br>1: None of the members suffer from major ailments<br>2: Any one member of the family member apart from the respondent suffers from either high blood pressure or diabetes | Categorical |
| Anxiety score | We use the four indicators capturing anxiety to calculate this score. Like in the other cases, we first normalize each of the four indicators to have zero mean and a standard deviation of one. Then we sum up the standardized measures and then again re-standardize such that the final score would have 0 mean and one standard deviation. | Continuous |
| State dummies | Three categories<br><br>1: West Bengal<br>2: Tamil Nadu<br>3: Other states | Categorical |

## 4. Results

*4.1 Descriptive statistics*

Table 3 gives an overview of the sample of respondents with respect to the individual compliance level during the survey through different indicators that we collected during the survey. The descriptive statistics on the associated indicators used to derive the perception about community compliance scores are also presented. Fifty six percent of our sample reported of maintaining social distance in the one week prior to the survey every time they interacted with others while over 82 percent report regular wearing of face-masks. Our respondents reported that they perceived 6 people out of 10 in their community on average maintained social distancing as well as stepped out only for essential purposes.

Table 3: Descriptive statistics of individual compliance and perceptions about community compliance behavior

| Individual compliance indicators (Proportions) | | | | |
|---|---|---|---|---|
| | Maintaining 6 ft. distance | Wearing face-masks | Using hand-sanitizers | Hand-washing |
| None of the times | 0.015 | 0.020 | 0.026 | 0.009 |
| Rarely | 0.044 | 0.013 | 0.062 | 0.020 |
| Half of the time | 0.089 | 0.024 | 0.088 | 0.059 |
| Most of the time | 0.296 | 0.123 | 0.234 | 0.213 |
| Always | 0.555 | 0.821 | 0.590 | 0.701 |
| **Perception about community compliance** | | | | |
| | | | Mean | Standard Deviation |
| *Indicators* | | | | |
| Out of 10 people in your community how many stepped only out for essential purpose | | | 6.098 | 2.730 |
| Out of 10 people in your community how many maintained 6 feet distance while talking | | | 6.261 | 2.667 |

The statistics for other variables used in the analysis is given in Table 4. We find that close to 46% of our sample respondents are females and majority of them are in the age cohort 30 to 50 years. Over 67% of them are educated at least at the graduate level. About 17% of them report of suffering from high blood-pressure or diabetes. As indicated earlier, we collected our data over lockdown phase 2 to lockdown phase 4 in India but 75% of our sample responded during lockdown phase 2.

Table 4: Descriptive Statistics

| **Variables** | **Proportion/ Mean** |
|---|---|
| *Outcome indicators* | |
| Always maintained 6 feet distance | 0.555 |
| Always used hand-sanitizer | 0.590 |
| Always used facemasks | 0.821 |
| Always washed hands | 0.701 |
| *Community compliance indicators* | |
| Community maintains social distancing (out of 10 in the community) | 6.261 |
| Community does not step out for non-essential purpose (Out of 10 in the community) | 6.098 |
| *Control variables* | |
| Community should not step out for non-essential purpose (Out of 10 in the community) | 7.604 |
| Community should maintain social distancing (out of 10 in the community) | 7.700 |
| *Lockdown phases* | |
| Phase 1 | 0.744 |
| Phase 2 | 0.201 |
| Phase 3 | 0.054 |
| *Age* | |
| Below 25 years | 0.195 |
| 26 to 30 years | 0.192 |
| 31 to 40 years | 0.272 |
| 41 to 50 years | 0.211 |
| Above 50 years | 0.129 |
| Female respondent | 0.459 |
| Marital status: Single | 0.646 |
| *Education* | |
| No formal education/ 10 years or below 10 years of education | 0.150 |
| Higher secondary/ College education | 0.176 |
| Graduate | 0.217 |
| Post graduate or above | 0.456 |
| *Occupation* | |
| Engaged as a regular/ contractual worker in government sector | 0.212 |
| Engaged as a contractual worker in government or private sector | 0.321 |
| Five or more members in the household | 0.370 |
| *Number of elderly* in the household in the age group 65 years and above | |
| No elderly | 0.571 |
| One elderly | 0.270 |
| More than one elderly | 0.159 |
| *Number of rooms* | |
| One or two rooms | 0.368 |
| Three rooms | 0.319 |
| More than three rooms | 0.313 |
| Lives in an owned standalone house | 0.403 |
| *Co-morbidity* | |
| Suffers from either high blood-pressure and diabetes | 0.172 |
| Anyone in the family suffers from either high blood-pressure and diabetes | 0.279 |
| *States* | |
| West Bengal | 0.336 |
| Tamil Nadu | 0.418 |
| Others | 0.246 |

*4.2 Regression Results*

First, we examine the relationship between the standardized individual compliance score and the perception about community compliance scores. Table 5 presents the regression results. We sequentially introduce different groups of covariates to confirm if the marginal effects of the community compliance remain stable with increasing explanatory power of the model. In the first specification, only the community compliance scores are incorporated in the model and in the second model, the lockdown dummy is included. In the next specifications, the household characteristics and then individual characteristics are included along with the lockdown dummy (a dummy to indicate phase 3 and phase 4 lockdown). In the final preferred specification, we put all the controls together with the state fixed effects (West Bengal, Tamil Nadu and others). Notably, we have controlled for perception about community beliefs about the compliance protocols which would control for any systematic bias that the respondent may have about the community. The findings indicate that the community compliance scores, in all the specification are statistically significant at 1% level. Notably, the effect size do not alter across specifications as we find one standard deviation increase in perception about community compliance abidance is associated with about 0.27-0.29 standard deviation increase in private compliance.

Of note is the fact that we find a significant negative effect of the subsequent lockdowns on compliance. In regression models where we replace the lockdown variable with the time since the first survey (in days), the findings remain similar as a robust negative association of individual compliance is found with this variable. This indicates the possibility of a fatigue effect along with non-compliance due to potential economic activity economic as discussed earlier.

Table 5: OLS regression estimates of individual compliance score on perception about community compliance

|  | (1) | (2) | (3) | (4) | (5) | (6) |
|---|---|---|---|---|---|---|
| Community compliance score | 0.287*** | 0.284*** | 0.280*** | 0.269*** | 0.271*** | 0.270*** |
|  | (0.029) | (0.029) | (0.029) | (0.033) | (0.033) | (0.033) |
| Phase3/ Phase 4 lockdown |  | -0.285*** |  | -0.226*** | -0.210*** |  |
|  |  | (0.074) |  | (0.071) | (0.071) |  |
| Time since the first survey (in days) |  |  |  |  |  | -0.012** |
|  |  |  |  |  |  | (0.005) |
| Household characteristics | No | No | Yes | Yes | Yes | Yes |
| Individual characteristics | No | No | No | Yes | Yes | Yes |
| State dummies | No | No | No | No | Yes | Yes |
| Observations | 934 | 934 | 920 | 919 | 919 | 919 |
| R-squared | 0.082 | 0.098 | 0.090 | 0.152 | 0.154 | 0.153 |

*Notes. The marginal effects from OLS regression with robust standard errors in parenthesis are reported. The dependent variable here is individual level standardized compliance score. *** p<0.01; * p<0.05; * p<0.1. Full regression is presented in appendix A1.*

Even though the marginal effects of community compliance remain largely stable in all the specifications, endogeneity from OVB and reverse causality remains a concern that prevents us from interpreting these results from the causal lens. With respect to OVB, there might be other unobservable factors correlated with community level compliance as well as his/ her own abidance that we did not account for. With respect to reverse causation, existing literature indicates that high association of bias in prevalence perception with one's own personal behavior. In particular, it found that people tend to perceive that their own behavior to be more common in the community than it actually is, often called the "false consensus effect" (Ross et al., 1977; Kuang et al. 2020). Therefore, because of the potential OVB and reverse causation, we may not be able to say higher community compliance is causally related to better abidance of the compliance protocols.

*4.3 Accounting for Omitted Variable Bias (OVB)*

To assess the potential changes in the coefficients after accounting for OVB, we use a strategy developed by Oster (2019) based on Altonji et al. (2005), based on the assumption that selection on unobservable variables can be gauged by the extent of selection on observables. For further elaboration, we consider the following regression equation:

$Y = \beta Z + \phi X + U$

Here $Y$ is the outcome variable, $Z$ is the primary variable of interest, $X$ is the vector of observable controls and $U$ is the set of all unobserved components. If $U$ is substantially high, our aim is to examine the corresponding changes in $\beta$. Here the primary assumption is:

$$\frac{\text{cov}(Z,U)}{\text{var}(U)} = \delta \frac{\text{cov}(Z,\beta X)}{Var(\beta X)}$$

This implies that the relationship between $Z$ and the unobserved component is proportional to the correlation between $Z$ and the observed component and $\delta$ gives the degree of proportionality. With this, it is possible to derive a consistent estimator of $\alpha$ as a function of two parameters: $\delta$ and $R_{max}$. $R_{max}$ is the $R$-squared value of a hypothetical regression which includes the observable as well as the unobserved components in the model. Among various options to compute the bound on $R_{max}$, based on data from a sample of randomized experiments, Oster (2016) proposes $R_{max} = 1.3 * R_0$, where $R_0$ is the $R$-squared value of the full model with observed covariates.[6] For parameter $\delta$, if we assume that the extent of bearing of the observables is at least as important as the effect of omitted variables, then the bound for $\delta \in [-1,1]$ (Altonji et al., 2005; Mukhopadhyay and Sahoo, 2016). Oster (2019) suggests two equivalent approaches to verify if the effect becomes indistinguishable from zero after controlling for the OVB. Firstly, this involves checking if $\delta$ for which the coefficient of interest turns 0 with $R_{max} = 1.3*R_0$ exceeds a threshold of 1. We also tested this with the $R_{max}$ value to be as high as $R_{max} = 2*R_0$. The argument is if $\delta$ exceeds 1 or -1, the extent of the association with unobservable has to be massively high, which may not be valid. Secondly, we test the coefficient value changes its sign if in the interval $\delta \in [-1,1]$ and $R_{max} = \pi*R_0$ where $\pi = 1.3 \ and \ \pi = 2$. If sign does not change, then even if the extent of association with the unobservable is as high as that with the observables and in either direction as well, the null hypothesis of $\beta = 0$ can be rejected. From Table 6, which presents the relevant results, we observe that $\delta$ is much higher than 1 (1.35-2.8). Similarly, we find $\beta$ to be in the range 0.27-0.28 when $\delta$ changes from 1 to -1 with $R_{max}$=0.195 ($R_{max} = 1.3*R_0$) and between 0.21 to 0.29 with $R_{max}$=0.3 ($R_{max} = 2*R_0$). This implies that the marginal effects of community compliance scores on community compliance on individual COVID-19 protocol abidance remains significant even after accounting for the potential OVB.

Table 6: Potential bias because of omitted variables

| | | Identified (estimated bias) | | | |
| --- | --- | --- | --- | --- | --- |
| | | $R_{max}$=0.195 | | $R_{max}$=0.3 | |
| Uncontrolled (R$^2$) | Controlled (R$^2$) | $\beta$ for $\delta \in [-1,1]$ | $\delta$ for $\beta = 0$ | $\beta$ for $\delta \in [-1,1]$ | $\delta$ for $\beta = 0$ |
| 0.280 | 0.271 | [0.277,0.262] | 2.79 | [0.286,0.209] | 1.35 |
| 0.081 | 0.154 | | | | |

*Notes.* The uncontrolled coefficient is from the regression of individual compliance score without any other control variables. The controlled coefficient is from its regression on the perception about community compliance, adjusting for all the control variables. We use the command *psacalc* in STATA 14.

---

[6] The bounding value of the cut off at 1.3 as per Oster (2016) allows effects from at-least 90% of the random experiments to survive.

Further, we apply other robustness checks as well to ensure that the relationship is robust and do not get alter with changes in specification, methods or the formulation of the dependent and independent variable.[7] These are as follows:

(i) **Individual compliance protocols:** We examine the relationship between abidance of the compliance protocol indicators separately with perception about community compliance for each the two protocol indicators we consider through a probit regression. The findings indicate positive relationship for most of these compliance indicators signifying that our results are robust (appendix table A2).

(ii) **With sub-samples:** Since most of the responses were from the states of West Bengal and Tamil Nadu, we ran the same regressions with respondents only from these two states. In a separate exercise, because majority of our respondents came from the bigger cities, we also run the same regressions separately for respondents from only the six biggest cities of India: New Delhi, Mumbai, Kolkata, Chennai, Hyderabad and Bangalore. In these regressions, we incorporate city level fixed effects that would control for city specific observables that include deaths, number of tests and administrative efforts among others that may be correlated with compliance protocol abidance. The inference from the regressions is qualitatively similar to what we observed with the full sample (appendix A3).

(iii) **Falsification tests:** For falsification test, we replace the outcome variable with a non-equivalent variable that is not expected to change in response to changes in the treatment. In our survey, we collect information on satisfaction about government measures taken on distribution of food ration to the poor and dealing with the internal migrant crisis that the country faced with the lockdown.[8] Since these variables are views on government actions which are not related to compliance or even COVID protocol awareness, they are unlikely to be related to the respondent's perception about community compliance. This is indeed found to be the case as we observe the coefficient of community compliance score is statistically indistinguishable from zero (appendix table A4).

*Relationship with lockdown and health conditions*

---

[7] All these results from robustness checks are given in appendix.
[8] Because of the sudden enforcement of the lockdown, there has been a mass exodus of migrant laborers from the cities to the villages leaving them starved and a number of deaths have been reported from across the country (Lancet, 2020)

In the first part of the analysis, we find the lockdown (phase 3 and 4) dummy is found to be negatively associated with individual compliance. Can higher perception about community compliance offset this decline in compliance over time? To answer this, we introduce an interaction of the community compliance variable and the lockdown dummy in the regression model and study the estimations (Table 4). We observe that the marginal effect of this interaction term to be positive implying an over-time increase in the potential effect of the respondent's perception about community compliance on her own compliance behavior. Hence, individual compliance which can potentially decline over time because of fatigue and economic conditions, enhancing perception on community level compliance can have a positive effect.

Studies have indicated individual with pre-existing chronic conditions like diabetes and high blood pressure have higher health risks (Fang et al. 2020; Klonoff and Umpierrez, 2020). Despite this, our regression results seem to indicate respondents with co-morbidities do not have significantly higher compliance (Table 7). Accordingly, we assess the heterogeneous effects of community compliance with individual compliance among these respondents (Table 4). The findings indicate that the community compliance is positively related and, statistically significant at 5% level.

We argue this finding has considerable policy implications. Unhealthy respondents who have higher health risk, can benefit from community level interventions that increase their perception about abidance, since it can have positive effect on their own compliance. Further, individual compliance which can potentially decline over time because of fatigue and economic conditions, enhancing perception on community level compliance can have a positive association with it, which can possibly be causal as well.

Table 7: Lockdown effects and effects on respondents with poor health

|  | Interaction with lockdown | Interaction with co-morbidity |
|---|---|---|
| *Ref. Lockdown (Phase 1)* |  |  |
| Lockdown (phase 3 and 4) | -0.202*** | -0.201*** |
|  | (0.071) | (0.071) |
| Community compliance score | 0.242*** | 0.241*** |
|  | (0.038) | (0.037) |
| Community compliance score*Lockdown (phase 3 and 4) | 0.142** |  |
|  | (0.066) |  |
| Suffering from either diabetes or high blood pressure | -0.111 | -0.121 |
|  | (0.117) | (0.116) |
| Community compliance score * Suffering from either diabetes or high blood pressure |  | 0.180*** |
|  |  | (0.066) |
| Household and individual characteristics | Yes | Yes |
| State dummies | Yes | Yes |
| Observations | 919 | 919 |
| R-squared | 0.157 | 0.159 |

*Notes. The marginal effects from OLS regression with robust standard errors in parenthesis are reported. The dependent variable here is individual level standardized compliance score. *** p<0.01; * p<0.05; * p<0.1. Full regression estimates are presented in appendix table A5.*

## 6. Discussion and policy implications

The main finding of the paper underscores that positive perceptions regarding community behavior can encourage individual COVID-19 compliance protocols. Additionally, affirmative perceptions regarding community are seen to sustain private compliance effort over time even with a potential fatigue effect and mobility restrictions that pose binding constraints on the income earning ability in general. Our work thus reaffirms the importance of social beliefs and norms on individual health focused behavior that literature has often focused on (Cialdini and Goldstein, 2004; Bicchieri, 2017; Van Bavel et al. 2020). Notably, literature on Ebola epidemics and even non-contagious ailments like HIV/ AIDS that documents outreach interventions facilitated through local community organizations led to increased awareness, contained the spread of the virus and lowered stigma (Gregson et al., 2013; Abramowitz et al. 2015; Pronyk et al. 2016). Community focused measures were recommended in other studies that have highlighted the role of norms in collectivistic cultures in predicting SARS-preventive behaviors (Chen and Ng, 2006).

The underlying implication of our results highlights the need for behavioral change interventions that use influential individuals from the local community to exert social pressure or incentives to sustain compliance. Social distancing measures and wearing masks are positive behaviors that should be made visible and salient so that it can stabilize a norm through visual cues. In this context, social media platforms can be used to engage peers and broadcast the use of masks and safety measures. If more people see their neighbors in their community wear masks and talk about staying at home during the pandemic, it might increase one's willingness to do the same. Such community level interventions may be more cost effective compared to individual level monitoring and vigilance and also less likely to raise mistrust and antagonism against the implementing authorities.

Importantly, our findings also signify the effectiveness of sensitization through community especially for individuals with pre-existing co-morbidities. As COVID infection propensity and mortality are known to be higher among people with co-morbidities (Ejaz et. al., 2020) community based interventions, in the light of our result, can be useful in reducing the fatality rate. Hence such interventions should be targeted more in the local public health centres or hospitals or even medicine shops where the likelihood of the co-morbid individuals visiting remains higher. Focused interventions related to community compliance through posters or periodic announcements in these places can be a good way to reinforce these motivations. While our paper identifies the significance of community interventions, it must be noted the design of such programs are largely contextual and constitutes an important future research agenda.

Along with these interventions that raise perception about community, targeted educational and communication messages periodically stressing the need for abiding by the pandemic protocol along with its relevance to the local community should be clearly stated. For longer term benefits, stronger institutions and policies that promote harmony and cooperation among the community members becomes pertinent. This becomes especially important in urban areas where community ties are relatively weaker. Further because our empirical assessment reveals that subsequent lockdown has reduced private compliance, imposing mobility restrictions that impede community interactions can potentially constrain individual compliance. While gauging the impact of deterrence measures like lockdowns on enhancing compliance and comparing that with other government measures is beyond the scope of the paper, we note that as a policy tool, lockdown may not serve to propagate voluntary protection effort and might even dampen it.

The findings should be interpreted with certain caveats and limitations. First, without exogenous variation in community compliance, we are unable to view the relationship between individual abidance of protocols and perception about community compliance from the causal lens. Though we account for the potential OVB, concerns about reverse causation remain valid. Secondly, snowball sampling and usage of social platforms raise concerns regarding the representativeness of the sample particularly if sections of the populations without smart phones are systematically excluded from the sample. Further, self-reported data on compliance, might also be associated with substantial upward bias. Nevertheless, given unprecedented nature of the outbreak and subsequent lockdowns, we feel these findings provide insight on an importance issue regarding compliance.

## 7. Conclusion

This paper examines the relationship between perceived community compliance and individual abidance to the COVID-19 compliance protocols. We find that community influence reinforcing individual compliance in data collected across India from April to May, 2020 during the lockdown. The heterogeneous effects indicate higher effects of the community over time and also on co-morbid individuals. Based on these findings, we highlight a number of policy recommendations based on using community networks to foster collective action that can raise compliance and arrest the growth of the infection. Apart from the policy implication, the paper assumes contextual importance as India is currently among those countries with the highest number of new infection and deaths.

## Appendix

Appendix 1:

In this section, we assume a functional form for the utility function, $U = U(x_i, h_i, I_i)$ as defined in section 2 and examine the theoretical predictions on some of empirical estimations. We make some modification to the health function and assume that health is a function of compliance $c_i$ and medical care $m_i$ for individual, $i$, both measured in time units. The basic results of the model, however, remain unchanged. The health function would then be

$$h = p(\bar{c}_i)c_i + (1 - p(\bar{c}_i))m_i \ldots\ldots (1)$$

Here, we define $p(\bar{c}_i)$ as the probability of staying healthy for a given perception about community health norm, $\bar{c}_i$. Further we define

$p(\bar{c}_i) = \theta + b\bar{c}_i, b > 0, 0 \leq \theta \leq 1 - b\bar{c}_i$........(2)

where $\theta$ is the health stock of the individual which is independent of community compliance and accounts for the pre-existing co-morbidities.

The time constraint is given by

$c_i + m_i + l_i = T$... (3), where $l_i$ is the work time and $T$ is the time endowment.

If the unit price of medical care is $p_m$, and $w$ is the wage, the income constraint is described as

$x_i + p_m m_i = wl_i = w(T - c_i - m_i)$...(3a)

The individual maximization problem, assuming additively separable utility function, can be stated as:

$$\max_{c_i, m_i} U(c_i, m_i) = [w(T - c_i - m_i) - p_m m_i] + [\theta + b\bar{c}_i]c_i + (1 - \theta - b\bar{c}_i)m_i - a(c_i - \bar{c})^2$$

$\frac{\partial U}{\partial c_i} = 0$, imply $-\frac{w}{2a} + \frac{\theta}{2a} + \bar{c}_i \left[\frac{b}{2a} + 1\right] = c_i$.....(4)

$\frac{\partial c_i}{\partial \bar{c}_i} = \left[\frac{b}{2a} + 1\right] > 0$......(5)

Hence, it is expected that with an increase in perception on community compliance, private compliance efforts would also increase.

Further, $\frac{\partial U}{\partial m_i} = 0$ gives $(1 - \theta - b\bar{c}_i) = (w + p_m)$ ......(6)

From (6) we have $\bar{c}_i = \frac{(1-\theta)-(w+p_m)}{b}$, substituting in (4) and differentiating with respect to $\theta$ we get

$$\frac{\partial c_i}{\partial \theta} = -\frac{1}{b} < 0$$

Thus, we find that people with lower health stock (hence with pre-existing co-morbidities) are expected to choose a higher compliance level given the probability of infection.

Table A1: OLS regression estimates of individual compliance score on perception about community compliance

| | (1) | (2) | (3) | (4) | (5) | (6) |
|---|---|---|---|---|---|---|
| Community compliance score | 0.287*** | 0.284*** | 0.280*** | 0.269*** | 0.271*** | 0.270*** |
| | (0.029) | (0.029) | (0.029) | (0.033) | (0.033) | (0.033) |
| Beliefs about community score | | | | -0.021 | -0.016 | -0.015 |
| | | | | (0.034) | (0.034) | (0.034) |
| Phase3/ Phase 4 lockdown | | | -0.285*** | -0.226*** | -0.210*** | |
| | | | (0.074) | (0.071) | (0.071) | |
| Time since the first survey (in days) | | | | | | -0.012** |
| | | | | | | (0.005) |
| *Age category (Ref. Below 25 years)* | | | | | | |
| 26 to 30 years | | | | -0.040 | -0.038 | -0.061 |
| | | | | (0.118) | (0.118) | (0.117) |
| 31 to 40 years | | | | 0.059 | 0.062 | 0.036 |
| | | | | (0.126) | (0.124) | (0.123) |
| 41 to 50 years | | | | 0.221 | 0.207 | 0.179 |
| | | | | (0.135) | (0.134) | (0.134) |
| Above 50 years | | | | 0.036 | 0.028 | 0.002 |
| | | | | (0.140) | (0.140) | (0.140) |
| *Sex (Ref. Male/ cannot say)* | | | | | | |
| Female | | | | 0.121* | 0.116 | 0.116* |
| | | | | (0.072) | (0.071) | (0.071) |
| *Marital Status (Ref. Cohabiting/ Married/ Others)* | | | | | | |
| Single | | | | 0.052 | 0.042 | 0.044 |
| | | | | (0.082) | (0.083) | (0.083) |
| *Education (Ref. Post graduate or above)* | | | | | | |
| No formal education/ 10 years or below 10 years of education | | | | -0.177 | -0.177 | -0.221* |
| | | | | (0.109) | (0.116) | (0.118) |
| Higher secondary/ College education | | | | 0.203** | 0.196** | 0.175** |
| | | | | (0.083) | (0.088) | (0.088) |
| Graduate | | | | 0.122 | 0.115 | 0.111 |
| | | | | (0.075) | (0.075) | (0.075) |
| Engaged as a regular/ contractual worker in government sector | | | | 0.275*** | 0.281*** | 0.299*** |
| | | | | (0.083) | (0.090) | (0.091) |
| Engaged as a contractual worker in government or private sector | | | | -0.062 | -0.059 | -0.050 |
| | | | | (0.094) | (0.095) | (0.094) |
| *Large family (Ref. Four members or less in the household)* | | | | | | |
| Five or more members in the household | | | -0.083 | -0.108 | -0.105 | -0.100 |
| | | | (0.073) | (0.076) | (0.077) | (0.078) |
| *Number of elderly (Ref. No elderly)* | | | | | | |
| One elderly in the household in the age group 65 years and above | | | 0.086 | 0.088 | 0.079 | 0.074 |
| | | | (0.075) | (0.076) | (0.076) | (0.076) |
| More than one elderly in the household in the age group 65 years and above | | | 0.081 | 0.123 | 0.112 | 0.105 |
| | | | (0.090) | (0.092) | (0.092) | (0.092) |
| *Number of rooms (Ref. One or two rooms)* | | | | | | |
| Three rooms | | | -0.114 | -0.045 | -0.047 | -0.049 |
| | | | (0.074) | (0.078) | (0.079) | (0.078) |
| More than three rooms | | | -0.028 | 0.003 | -0.007 | -0.008 |
| | | | (0.081) | (0.087) | (0.086) | (0.087) |
| *Type of residence (Ref. Lives in a rented house or apartment complex)* | | | | | | |
| Lives in an owned standalone house | | | | 0.144** | 0.106 | 0.109 |
| | | | | (0.070) | (0.074) | (0.074) |
| Co-morbidity | | | | | -0.125 | -0.122 |
| | | | | | (0.117) | (0.117) |
| Family co-morbidity | | | | | -0.145 | -0.158 |
| | | | | | (0.098) | (0.097) |
| Anxiety score | | | | | 0.025* | 0.024* |
| | | | | | (0.013) | (0.013) |
| *State dummies (Ref. West Bengal)* | | | | | | |
| Tamil Nadu | | | | | | -0.061 | -0.062 |
| | | | | | | (0.095) | (0.096) |
| Others | | | | | | -0.120 | -0.109 |
| | | | | | | (0.087) | (0.090) |
| Constant | 0.000 | 0.073** | -0.013 | -0.133 | -0.061 | 0.001 |
| | (0.031) | (0.036) | (0.070) | (0.112) | (0.122) | (0.126) |
| Observations | 934 | 934 | 920 | 919 | 919 | 919 |
| R-squared | 0.082 | 0.098 | 0.090 | 0.152 | 0.154 | 0.153 |

*Notes. The marginal effects from OLS regression with robust standard errors in parenthesis are reported. The dependent variable here is individual level standardized compliance score. *** $p<0.01$; ** $p<0.05$; * $p<0.1$.*

Table A2: OLS regression estimates of individual compliance indicators on perception about community compliance

|  | Maintaining social distancing) | Wearing face-masks | Using hand-sanitizers | Hand-washing |
|---|---|---|---|---|
| Community maintaining social-distancing | 0.308*** | 0.158*** | 0.258*** | 0.226*** |
|  | (0.040) | (0.040) | (0.041) | (0.040) |
| Community not going out for non-essential purpose | 0.069** | 0.030 | 0.037 | 0.078*** |
|  | (0.032) | (0.032) | (0.034) | (0.029) |
| Beliefs about community score | -0.107*** | -0.023 | -0.010 | -0.057 |
|  | (0.037) | (0.037) | (0.042) | (0.039) |
| Phase3/ Phase 4 lockdown | -0.222*** | -0.108 | -0.067 | -0.241*** |
|  | (0.072) | (0.074) | (0.077) | (0.080) |
| *Age category (Ref. Below 25 years)* | | | | |
| 26 to 30 years | 0.057 | -0.191 | -0.000 | 0.039 |
|  | (0.115) | (0.123) | (0.115) | (0.109) |
| 31 to 40 years | 0.200 | 0.065 | -0.069 | 0.024 |
|  | (0.127) | (0.130) | (0.126) | (0.128) |
| 41 to 50 years | 0.356*** | 0.099 | 0.011 | 0.141 |
|  | (0.132) | (0.137) | (0.131) | (0.135) |
| Above 50 years | 0.293** | 0.012 | -0.097 | -0.029 |
|  | (0.140) | (0.146) | (0.144) | (0.149) |
| *Sex (Ref. Male/ cannot say)* | | | | |
| Female | 0.128* | -0.018 | 0.097 | 0.058 |
|  | (0.069) | (0.074) | (0.074) | (0.074) |
| *Marital Status (Ref. Cohabiting/ Married/ Others)* | | | | |
| Single | -0.046 | 0.029 | 0.096 | 0.018 |
|  | (0.084) | (0.093) | (0.082) | (0.087) |
| *Education (Ref. Post graduate or above)* | | | | |
| No formal education/ 10 years or below 10 years of education | -0.271** | -0.176 | 0.087 | -0.265** |
|  | (0.122) | (0.108) | (0.115) | (0.116) |
| Higher secondary/ College education | 0.065 | 0.109 | 0.234** | 0.093 |
|  | (0.097) | (0.095) | (0.091) | (0.085) |
| Graduate | 0.054 | 0.080 | 0.167* | 0.073 |
|  | (0.081) | (0.077) | (0.086) | (0.076) |
| Engaged as a regular/ contractual worker in government sector | 0.150 | 0.319*** | 0.201** | 0.217** |
|  | (0.094) | (0.090) | (0.097) | (0.085) |
| Engaged as a contractual worker in government or private sector | -0.145* | 0.112 | -0.074 | -0.106 |

|  | (0.087) | (0.104) | (0.088) | (0.089) |
|---|---|---|---|---|
| *Large family (Ref. Four members or less in the household)* | | | | |
| Five or more members in the household | -0.025 | -0.059 | -0.107 | -0.119 |
|  | (0.076) | (0.077) | (0.076) | (0.080) |
| *Number of elderly (Ref. No elderly)* | | | | |
| One elderly in the household in the age group 65 years and above | -0.047 | 0.066 | 0.121 | 0.089 |
|  | (0.078) | (0.078) | (0.079) | (0.081) |
| More than one elderly in the household in the age group 65 years and above | -0.054 | 0.083 | 0.120 | 0.124 |
|  | (0.101) | (0.089) | (0.097) | (0.094) |
| *Number of rooms (Ref. One or two rooms)* | | | | |
| Three rooms | -0.053 | -0.064 | -0.064 | 0.050 |
|  | (0.081) | (0.075) | (0.082) | (0.085) |
| More than three rooms | 0.004 | -0.189** | -0.021 | 0.162* |
|  | (0.091) | (0.092) | (0.093) | (0.085) |
| *Type of residence (Ref. Lives in a rented house or apartment complex)* | | | | |
| Lives in an owned standalone house | 0.117 | 0.037 | 0.049 | 0.143** |
|  | (0.074) | (0.077) | (0.074) | (0.070) |
| Co-morbidity | -0.146 | -0.045 | -0.002 | -0.173 |
|  | (0.123) | (0.102) | (0.108) | (0.128) |
| Family co-morbidity | -0.175* | -0.076 | -0.114 | -0.049 |
|  | (0.100) | (0.087) | (0.088) | (0.103) |
| Anxiety score | 0.015 | 0.033** | 0.005 | 0.005 |
|  | (0.012) | (0.013) | (0.011) | (0.012) |
| *State dummies (Ref. West Bengal)* | | | | |
| Tamil Nadu | 0.181* | -0.176* | -0.182* | -0.091 |
|  | (0.105) | (0.096) | (0.108) | (0.096) |
| Others | 0.048 | -0.227** | -0.059 | -0.052 |
|  | (0.091) | (0.102) | (0.097) | (0.087) |
| Constant | -0.130 | 0.117 | -0.050 | -0.048 |
|  | (0.151) | (0.122) | (0.139) | (0.124) |
| Observations | 919 | 919 | 919 | 919 |
| R-squared | 0.157 | 0.088 | 0.113 | 0.128 |

*Notes. The marginal effects from OLS regression with robust standard errors in parenthesis are reported. The dependent variable here is individual level standardized compliance score. *** p<0.01; * p<0.05; * p<0.1.*

Table A3: OLS regression estimates of individual compliance on perception about community compliance for West Bengal and Tamil Nadu and metropolitan cities

|  | West Bengal and Tamil Nadu only | Metropolitan cities |
|---|---|---|
| Community compliance score | 0.256*** | 0.211*** |
|  | (0.036) | (0.056) |
| Beliefs about community score | -0.001 | -0.058 |
|  | (0.039) | (0.054) |
| Phase3/ Phase 4 lockdown | -0.312*** | -0.142 |
|  | (0.084) | (0.121) |
| *Age category (Ref. Below 25 years)* |  |  |
| 26 to 30 years | 0.087 | -0.035 |
|  | (0.122) | (0.174) |
| 31 to 40 years | 0.205 | 0.082 |
|  | (0.142) | (0.163) |
| 41 to 50 years | 0.272* | 0.125 |
|  | (0.154) | (0.189) |
| Above 50 years | 0.119 | -0.257 |
|  | (0.158) | (0.199) |
| *Sex (Ref. Male/ cannot say)* |  |  |
| Female | 0.103 | -0.030 |
|  | (0.082) | (0.102) |
| *Marital Status (Ref. Cohabiting/ Married/ Others)* |  |  |
| Single | 0.001 | -0.023 |
|  | (0.086) | (0.120) |
| *Education (Ref. Post graduate or above)* |  |  |
| No formal education/ 10 years or below 10 years of education | -0.179 | 0.509 |
|  | (0.122) | (0.324) |
| Higher secondary/ College education | 0.199** | -0.021 |
|  | (0.100) | (0.172) |
| Graduate | 0.116 | 0.219** |
|  | (0.086) | (0.092) |
| Engaged as a regular/ contractual worker in government sector | 0.299*** | -0.027 |
|  | (0.099) | (0.164) |
| Engaged as a contractual worker in government or private sector | -0.252** | -0.204* |
|  | (0.114) | (0.114) |
| *Large family (Ref. Four members or less in the household)* |  |  |
| Five or more members in the household | -0.067 | -0.248* |
|  | (0.084) | (0.126) |
| *Number of elderly (Ref. No elderly)* |  |  |
| One elderly in the household in the age group 65 years and above | 0.126 | 0.061 |
|  | (0.083) | (0.112) |
| More than one elderly in the household in the age group 65 years and above | 0.196* | 0.205 |
|  | (0.105) | (0.125) |
| *Number of rooms (Ref. One or two rooms)* |  |  |
| Three rooms | -0.005 | 0.074 |
|  | (0.093) | (0.115) |
| More than three rooms | 0.028 | 0.144 |
|  | (0.094) | (0.131) |
| *Type of residence (Ref. Lives in a rented house or apartment* |  |  |
| *complex)* | | |
| Lives in an owned standalone house | 0.129 | 0.182* |
| | (0.081) | (0.108) |
| Co-morbidity | -0.121 | 0.062 |
| | (0.124) | (0.165) |
| Family co-morbidity | -0.172* | -0.279* |
| | (0.098) | (0.150) |
| Anxiety score | 0.032** | 0.027 |
| | (0.014) | (0.022) |
| *State dummies (Ref. West Bengal)* | | |
| Tamil Nadu | -0.080 | |
| | (0.099) | |
| *City (Ref. Bangalore)* | | |
| Chennai | | -0.397 |
| | | (0.265) |
| Delhi | | -0.011 |
| | | (0.209) |
| Hyderabad | | -0.406 |
| | | (0.264) |
| Kolkata | | -0.095 |
| | | (0.200) |
| Mumbai | | -0.179 |
| | | (0.240) |
| Constant | -0.105 | 0.178 |
| | (0.137) | (0.253) |
| Observations | 692 | 377 |
| R-squared | 0.189 | 0.162 |

*Notes. The marginal effects from OLS regression with robust standard errors in parenthesis are reported. The dependent variable here is individual level standardized compliance score. \*\*\* p<0.01; \* p<0.05; \* p<0.1.*

Table A4: Falsification Tests

|  | Migrant Labour | Distribution of food | Safety gears for health workers |
|---|---|---|---|
| Community compliance score | 0.070** | 0.012 | -0.007 |
|  | (0.035) | (0.037) | (0.036) |
| Beliefs about community score | 0.063* | 0.112*** | 0.094*** |
|  | (0.036) | (0.038) | (0.035) |
| Phase3/ Phase 4 lockdown | 0.045 | 0.075 | 0.004 |
|  | (0.077) | (0.080) | (0.079) |
| *Age category (Ref. Below 25 years)* |  |  |  |
| 26 to 30 years | -0.278*** | -0.231** | -0.174 |
|  | (0.104) | (0.108) | (0.106) |
| 31 to 40 years | -0.347*** | -0.291** | -0.327*** |
|  | (0.114) | (0.116) | (0.115) |
| 41 to 50 years | -0.294** | -0.339*** | -0.342*** |
|  | (0.121) | (0.123) | (0.123) |
| Above 50 years | -0.136 | -0.216 | -0.161 |
|  | (0.139) | (0.141) | (0.138) |
| *Sex (Ref. Male/ cannot say)* |  |  |  |
| Female | 0.050 | 0.042 | -0.005 |
|  | (0.066) | (0.070) | (0.066) |
| *Marital Status (Ref. Cohabiting/ Married/ Others)* |  |  |  |
| Single | 0.054 | 0.063 | 0.085 |
|  | (0.080) | (0.081) | (0.079) |
| *Education (Ref. Post graduate or above)* |  |  |  |
| No formal education/ 10 years or below 10 years of education | 0.556*** | 0.451*** | 0.518*** |
|  | (0.121) | (0.128) | (0.122) |
| Higher secondary/ College education | 0.001 | -0.055 | -0.113 |
|  | (0.096) | (0.101) | (0.097) |
| Graduate | 0.137* | 0.168** | 0.180** |
|  | (0.075) | (0.077) | (0.075) |
| Engaged as a regular/ contractual worker in government sector | 0.232** | 0.234** | 0.265*** |
|  | (0.093) | (0.096) | (0.091) |
| Engaged as a contractual worker in government or private sector | 0.075 | -0.015 | -0.003 |
|  | (0.073) | (0.078) | (0.074) |
| *Large family (Ref. Four members or less in the household)* |  |  |  |
| Five or more members in the household | 0.061 | 0.157** | 0.143** |
|  | (0.069) | (0.074) | (0.072) |
| *Number of elderly (Ref. No elderly)* |  |  |  |
| One elderly in the household in the age group 65 years and above | -0.173** | -0.187** | -0.228*** |
|  | (0.071) | (0.077) | (0.074) |
| More than one elderly in the household in the age group 65 years and above | 0.021 | 0.019 | -0.118 |
|  | (0.090) | (0.096) | (0.092) |
| *Number of rooms (Ref. One or two rooms)* |  |  |  |
| Three rooms | -0.177** | -0.106 | -0.193** |
|  | (0.076) | (0.081) | (0.079) |
| More than three rooms | -0.013 | -0.013 | -0.105 |
|  | (0.084) | (0.089) | (0.083) |
| *Type of residence (Ref. Lives in a rented house or apartment complex)* | | | |
| Lives in an owned standalone house | -0.026 | -0.054 | -0.072 |
| | (0.066) | (0.069) | (0.067) |
| Co-morbidity | -0.213* | -0.134 | -0.062 |
| | (0.109) | (0.117) | (0.111) |
| Family co-morbidity | 0.104 | 0.111 | 0.021 |
| | (0.090) | (0.095) | (0.089) |
| Anxiety score | 0.010 | 0.015 | 0.031** |
| | (0.012) | (0.013) | (0.013) |
| *State dummies (Ref. West Bengal)* | | | |
| Tamil Nadu | 0.543*** | 0.217** | 0.372*** |
| | (0.099) | (0.102) | (0.101) |
| Others | 0.111 | 0.045 | 0.187** |
| | (0.077) | (0.081) | (0.079) |
| Constant | -0.178 | -0.054 | -0.032 |
| | (0.122) | (0.132) | (0.127) |
| | | | |
| Observations | 919 | 919 | 919 |
| R-squared | 0.260 | 0.160 | 0.219 |

*Notes. The marginal effects from OLS regression with robust standard errors in parenthesis are reported. The dependent variable here is individual level standardized compliance score. \*\*\* $p<0.01$; \* $p<0.05$; \* $p<0.1$.*

Table A5: Interaction effects

|  | (1) | (2) |
|---|---|---|
| Community compliance score | 0.242*** | 0.241*** |
|  | (0.038) | (0.037) |
| Phase3/ Phase 4 lockdown | -0.202*** | -0.201*** |
|  | (0.071) | (0.071) |
| Community compliance score* Phase3/ Phase 4 lockdown | 0.142** |  |
|  | (0.066) |  |
| Co-morbidity | -0.111 | -0.121 |
|  | (0.117) | (0.116) |
| Community compliance score* Co-morbidity |  | 0.180*** |
|  |  | (0.066) |
| Beliefs about community score | -0.015 | -0.013 |
|  | (0.034) | (0.034) |
| *Age category (Ref. Below 25 years)* |  |  |
| 26 to 30 years | -0.038 | -0.043 |
|  | (0.118) | (0.118) |
| 31 to 40 years | 0.053 | 0.055 |
|  | (0.124) | (0.124) |
| 41 to 50 years | 0.193 | 0.194 |
|  | (0.134) | (0.135) |
| Above 50 years | 0.023 | 0.013 |
|  | (0.140) | (0.140) |
| *Sex (Ref. Male/ cannot say)* |  |  |
| Female | 0.118* | 0.118* |
|  | (0.070) | (0.070) |
| *Marital Status (Ref. Cohabiting/ Married/ Others)* |  |  |
| Single | 0.036 | 0.047 |
|  | (0.083) | (0.083) |
| *Education (Ref. Post graduate or above)* |  |  |
| No formal education/ 10 years or below 10 years of education | -0.168 | -0.183 |
|  | (0.115) | (0.116) |
| Higher secondary/ College education | 0.184** | 0.203** |
|  | (0.088) | (0.088) |
| Graduate | 0.118 | 0.114 |
|  | (0.075) | (0.075) |
| Engaged as a regular/ contractual worker in government sector | 0.285*** | 0.285*** |
|  | (0.090) | (0.090) |
| Engaged as a contractual worker in government or private sector | -0.054 | -0.042 |
|  | (0.095) | (0.095) |
| *Large family (Ref. Four members or less in the household)* |  |  |
| Five or more members in the household | -0.109 | -0.102 |
|  | (0.077) | (0.077) |
| *Number of elderly (Ref. No elderly)* |  |  |
| One elderly in the household in the age group 65 years and above | 0.077 | 0.075 |
|  | (0.076) | (0.076) |
| More than one elderly in the household in the age group 65 years and above | 0.107 | 0.108 |
|  | (0.092) | (0.092) |
| *Number of rooms (Ref. One or two rooms)* |  |  |
| Three rooms | -0.049 | -0.045 |
|  | (0.078) | (0.078) |
| More than three rooms | -0.016 | -0.015 |
|  | (0.086) | (0.086) |
| *Type of residence (Ref. Lives in a rented house or apartment complex)* | | |
| Lives in an owned standalone house | 0.101 | 0.107 |
| | (0.075) | (0.074) |
| Family co-morbidity | -0.155 | -0.166* |
| | (0.097) | (0.097) |
| Anxiety score | 0.023* | 0.025* |
| | (0.014) | (0.013) |
| *State dummies (Ref. West Bengal)* | | |
| Tamil Nadu | -0.070 | -0.061 |
| | (0.096) | (0.096) |
| Others | -0.115 | -0.132 |
| | (0.088) | (0.088) |
| Constant | -0.046 | -0.059 |
| | (0.123) | (0.122) |
| | | |
| Observations | 919 | 919 |
| R-squared | 0.157 | 0.159 |

*Notes. The marginal effects from OLS regression with robust standard errors in parenthesis are reported. The dependent variable here is individual level standardized compliance score. \*\*\* p<0.01; \* p<0.05; \* p<0.1.*